\def\tc{t_{\rm cool}}
\def\tff{t_{\rm ff}}
\def\tctff{\tc / \tff}

\documentclass{emulateapj}

\begin{document}

\slugcomment{\apj\ Letters, submitted 20 Feb 2015 (Printed \today)}
\title{Supernova Sweeping and Black-Hole Feedback in Elliptical Galaxies}
\author{G. Mark Voit\altaffilmark{1,2}, Megan Donahue\altaffilmark{1}, Brian W. O'Shea\altaffilmark{1}, Greg L. Bryan\altaffilmark{3}, Ming Sun\altaffilmark{4}, Norbert Werner\altaffilmark{5} 
         } 
\altaffiltext{1}{Department of Physics and Astronomy,
                 Michigan State University,
                 East Lansing, MI 48824} 
\altaffiltext{2}{voit@pa.msu.edu}
\altaffiltext{3}{Department of Astronomy,
                 Columbia University,
                 New York, NY} 
\altaffiltext{4}{Department of Physics and Astronomy,
                 University of Alabama Huntsville,
                 Huntsville, AL} 
\altaffiltext{5}{Department of Physics,
                 Stanford University,
                 Palo Alto, CA} 
                           
\begin{abstract}
Most of the massive elliptical galaxies in the universe stopped forming stars billions of years ago, even though plenty of hot gas remains available for star formation.  Here we present compelling evidence indicating that quenching of star formation depends on both black-hole feedback and Type Ia supernova heating.  We analyze {\em Chandra} X-ray observations of ten massive ellipticals, five with extended, potentially star-forming multiphase gas and five single-phase ellipticals with no star formation.  The ratio of cooling time to freefall time at 1--10~kpc in the multiphase galaxies is $\tc/\tff \approx 10$, indicating that precipitation-driven feedback limits cooling but does not eliminate condensation.  In the same region of the single-phase galaxies, the radial profiles of gas entropy are consistent with a thermally stable ($\tc/\tff > 20$) supernova-driven outflow that sweeps stellar ejecta out of the galaxy.  However, in one of those single-phase ellipticals (NGC~4261) we find $\tc/\tff \lesssim 10$ at $< 300$~pc.  Notably, its jets are $\sim 50$ times more powerful than in the other nine ellipticals, in agreement with models indicating that precipitation near the black hole should switch its fueling mode from Bondi-like accretion to cold chaotic accretion.  We conclude by hypothesizing that particularly strong black-hole outbursts can shut off star formation in massive elliptical galaxies by boosting the entropy of the hot gas and flipping the system into the supernova-sweeping state.
\end{abstract}

\keywords{galaxies: clusters: intracluster medium}

\section{Introduction}
\label{sec-Intro}

\setcounter{footnote}{0}

Abundant circumstantial evidence supports the hypothesis that black-hole feedback suppresses star formation in massive galaxies \citep[e.g.,][]{mn07,McNamaraNulsen2012NJPh...14e5023M} but a deep mystery remains: How does accretion fueling of the black-hole engine become closely coupled with the vast hot-gas atmosphere surrounding it?   Precipitation-driven feedback models are providing particularly promising answers to this question.  

Numerical simulations show that cool clouds can precipitate out of a galaxy's hot-gas atmosphere via thermal instability if it is in a state of global thermal balance, with heating approximately equal to cooling \citep[][see also \citet{ps05}]{McCourt+2012MNRAS.419.3319M,Sharma+2012MNRAS.420.3174S,Gaspari+2012ApJ...746...94G}. The critical criterion for precipitation depends on the ratio between the time $\tc$ required for gas at temperature $T$ to radiate $3kT/2$ per particle and the free-fall time $\tff = (2r/g)^{1/2}$ required for a dense cool cloud to fall from a radius $r$ at the local gravitational acceleration $g$.  When the average $\tctff$ ratio is less than 10, then cooling is fast enough for some of the hot gas to condense into cold clouds and precipitate out of the hot medium. 

Precipitation itself plays an essential role in maintaining the required state of global thermal balance because it provides fuel for accretion.  Recent numerical simulations of this feedback loop have been very encouraging, because they show that ``chaotic cold accretion" of precipitating clouds can produce a black-hole fueling rate two orders of magnitude greater than the Bondi rate and can therefore generate a feedback response that brings the system into approximate balance at $\tctff \approx 10$ \citep[][see also \citet{ps10}]{Gaspari+2013MNRAS.432.3401G,Gaspari+2014arXiv1407.7531G,LiBryan2014ApJ...789...54L,LiBryan2014ApJ...789..153L}.

We have recently presented measurements of galaxy-cluster atmospheres that strongly support this picture.  The amount of multiphase gas in the cores of galaxy clusters steeply anticorrelates with $\min( \tctff)$, indicating that the precipitation rate, and therefore the feedback response, depends very strongly on this parameter \citep{VoitDonahue2015ApJ...799L...1V}.  Even more compelling is the finding that $\tctff \approx 10$ is a global lower limit on this ratio at all radii and at all temperatures among galaxy clusters in the {\em ACCEPT}\footnote{ \citet{Cavagnolo+09}, http://www.pa.msu.edu/astro/MC2/accept/ } database \citep{Voit+2014arXiv1409.1598V}.   

Here we show, using {\em Chandra} observations originally presented by \citet{Werner+2012MNRAS.425.2731W,Werner+2014MNRAS.439.2291W}, that the $\tctff \approx 10$ precipitation limit also applies to massive elliptical galaxies. Those observations revealed a close link between the presence of extended multiphase gas in ellipticals and the thermodynamic properties of their hot-gas atmospheres.  This paper interprets that finding in terms of a model combining supernova-driven outflows and precipitation-driven feedback. Section~2 briefly summarizes the key features of the Werner et al. ellipticals.  Section~3 shows that Type Ia supernovae (SNIa) can drive outflows that sweep stellar ejecta out of ellipticals with velocity dispersion $\sigma_v \gtrsim 250 \, {\rm km \, s^{-1}}$ but cannot prevent precipitation within the central kiloparsec.  Section~4 argues that outbursts of precipitation-driven black-hole feedback at $\lesssim 1$~kpc prevent $\tc$ from falling much below $10 \tff$. Section~5 shows that the black hole's jets are poorly coupled to the 1--10 kpc region but usually manage to maintain an isentropic core at $\lesssim 0.5$~kpc.  Section~6 concludes by hypothesizing that black-hole feedback shuts off star formation in massive ellipticals by switching on supernova sweeping.

\section{Radial Profiles \& Precipitation}
\label{sec-Profiles}

Profiles of electron density ($n_e$), temperature ($T$), and entropy index ($K \equiv kTn_e^{-2/3}$) for ten massive elliptical galaxies from \citet{Werner+2012MNRAS.425.2731W,Werner+2014MNRAS.439.2291W} are shown in Figure~1 as functions of radius $r = (1 \, {\rm kpc}) r_{\rm kpc} $.  Dashed lines and blue symbols show the subset of five with multiphase emission-line nebulae at $r \sim 1$--10 kpc. Solid lines and red or purple symbols show the subset of five without extended nebulae. The density panel shows that $n_e \approx (6 \times 10^{-2} \, {\rm cm^{-3}}) r_{\rm kpc}^{-1.3}$ is a good approximation at 1--10~kpc for the single-phase ellipticals, whereas the multiphase ellipticals tend to have density profiles closer to $n_e \propto r^{-1}$.  The temperature panel shows that the multiphase galaxies are systematically cooler than those without extended nebulae. In this sample, the temperature difference corresponds to a difference in velocity dispersion: The five galaxies with extended nebulae have one-dimensional velocity dispersions in the range $\sigma_v = 217$--$255 \, {\rm km \, s^{-1}}$, and the five without extended nebulae have $\sigma_v = 263$--$336 \, {\rm km \, s^{-1}}$, according to the HyperLeda database.\footnote{http://leda.univ-lyon1.fr}

\begin{figure*}[t]
\includegraphics[width=5.8in, trim = -1.5in 0.0in 0.0in 0.0in]{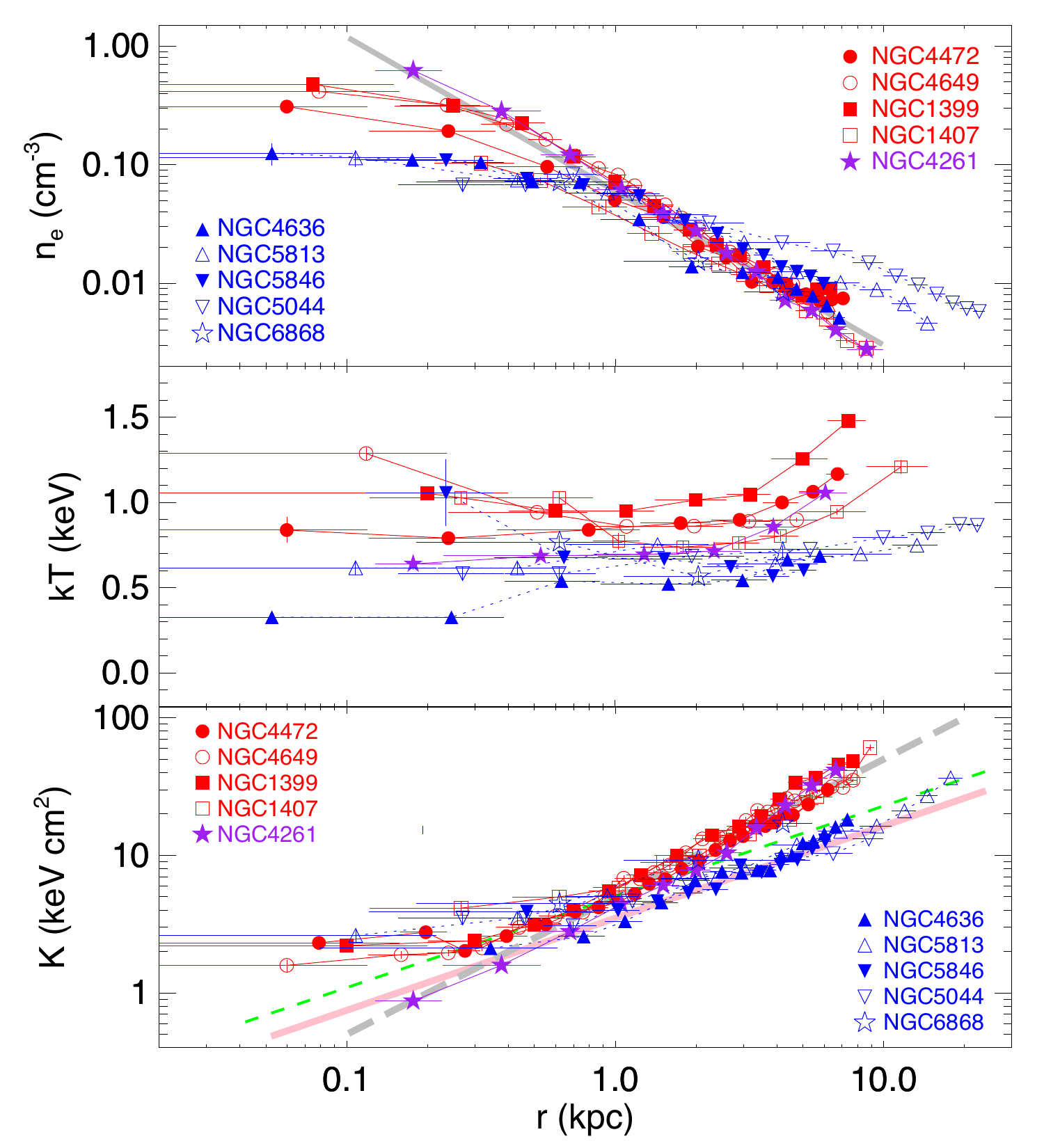} \\
\caption{ \footnotesize 
Radial profiles of electron density $n_e$ (top panel), gas temperature $T$ (middle panel), and entropy index $K \equiv kTn_e^{-2/3}$ (bottom panel) in massive ellipticals with extended emission-line nebulae (dotted lines, blue symbols) and without extended nebulae (solid lines, red or purple symbols).  A solid grey line in the top panel shows $n_e(r) = (6 \times 10^{-2} \, {\rm cm^{-3}}) r_{\rm kpc}^{-1.3}$.  A long-dashed grey line in the bottom panel shows $K(r) = (5 \, {\rm keV \, cm^2}) r_{\rm kpc}$.  A solid pink line in the bottom panel shows $K(r) = (3.5 \, {\rm keV \, cm^2}) r_{\rm kpc}^{2/3}$, which corresponds to the precipitation threshold at $t_{\rm cool} / t_{\rm ff} \approx 10$ for galaxies in this mass range. A short-dashed green line shows $K(r) = (5 \, {\rm keV \, cm^2}) r_{\rm kpc}^{2/3}$, which corresponds to the unstable locus at which SNIa heating balances radiative cooling.
\vspace*{1em}
\label{Werner_ellipticals}}
\end{figure*}

Combining $n_e(r)$ and $T(r)$ to obtain $K(r)$ reveals that the hot atmospheres of multiphase ellipticals distinctly differ from those without extended nebulae.  The single-phase elliptical galaxies follow the power law $K(r) \approx (5 \, {\rm keV \, cm^2}) r_{\rm kpc}$ in the 1--10~kpc range, whereas the multiphase elliptical galaxies tend to follow $K(r) \approx (3.5 \, {\rm keV \, cm^2}) r_{\rm kpc}^{2/3}$.
  
This latter power law corresponds to the precipitation threshold predicted by \citet{Sharma+2012MNRAS.427.1219S} and confirmed in central cluster galaxies by \citet{Voit+2014arXiv1409.1598V}.  Setting $t_{\rm cool} = 10 t_{\rm ff} \approx 10 r \sigma_v^{-1}$ yields
\begin{equation}
   K_{\rm precip}(r) \approx (3.5 \, {\rm keV \, cm^2}) \, T_{\rm keV}^{1/3} \, \Lambda_{\rm 3e-23}^{2/3} \,  
   			\sigma_{250}^{-2/3} \, r_{\rm kpc}^{2/3} \; \; ,
\end{equation}
where $\Lambda = (3 \times 10^{-23} \, {\rm erg \, cm^3 \, s^{-1}}) \Lambda_{\rm 3e-23}$ is the usual radiative cooling function for gas in collisional ionization equilibrium and depends on elemental abundances in this temperature range, with $\Lambda_{\rm 3e-23} \approx 1$ for a solar-abundance plasma.  Just as in central cluster galaxies, the entropy profiles of elliptical galaxies with extended multiphase gas at 1--10~kpc appear to be maintained by precipitation-driven feedback, strengthening the overall case for this mechanism.  However, something else must be happening at these radii in single-phase elliptical galaxies.

\section{Supernova Sweeping}
\label{sec-Sweeping}

We propose that the difference between multiphase and single-phase ellipticals arises because SNIa are successfully sweeping gas ejected by the old stellar population out of the single-phase ellipticals.  Stars belonging to an old stellar population shed matter with a specific mass ejection rate
\begin{equation}
  \alpha = \alpha_{-19} \times 10^{-19} \, {\rm s}^{-1} \approx ( 3 \times 10^{-12} \, {\rm yr^{-1}}) \, \alpha_{-19} \; \; ,
\end{equation}
where $\alpha_{-19}$ is within a factor of 2 of unity but depends somewhat on the age of the stellar population and the low-mass tail of the stellar mass function \citep[e.g.,][]{Mathews1990ApJ...354..468M,Conroy2014arXiv1406.3026C}.  The ejecta cannot linger, because otherwise the galaxy would either contain more gas or would be forming stars at a rate equivalent to the mass ejection rate \citep[e.g.,][]{MathewsBrighenti2003ARAA..41..191M,VoitDonahue2011ApJ...738L..24V}.  Instead, the time-averaged outflow rate of stellar ejecta at radius $r$ must be
\begin{equation}
 \dot{M}(r) \approx (0.09 \, M_\odot \, {\rm yr^{-1}} ) \, \alpha_{-19} \, \sigma_{250}^2 \, r_{\rm kpc} \, \, , 
\end{equation}
where we have assumed that the stellar mass-density profile is approximately $\propto r^{-2}$ with a one-dimensional velocity dispersion $\sigma_v = (250 \, {\rm km \, s^{-1}}) \sigma_{250}$, for which the stellar mass within radius $r$ is $\approx ( 1.4 \times 10^{10} \, M_\odot ) \sigma_{250}^2 r_{\rm kpc}$.  Given the power-law electron density profile shown in Figure~1, the average outflow velocity is then
\begin{equation}
 v  \approx (5 \, {\rm km \, s^{-1}} )  \, \alpha_{-19} \, \sigma_{250}^{2} \, r_{\rm kpc}^{0.3}   \, \, , 
\end{equation}
implying that a steady outflow would be subsonic and therefore close to hydrostatic equilibrium.

In order to illustrate how supernovae can produce an outflow with the power-law entropy profile observed in single-phase ellipticals, let us temporarily ignore radiative cooling.  In that case the entropy equation for steadily outflowing gas can be expressed as
\begin{equation}
  \frac {d \ln K} {dt} = \frac {2} {3} \frac {H} {nkT} - \frac {5} {3} \frac {\alpha \rho_*} {\mu m_p n}  \, \, ,
\end{equation}
where $H$ is the heating rate per unit volume, $n$ is the number density of gas particles, $\mu m_p$ is the mean mass per gas particle, and $\rho_*$ is the local stellar mass density.  The first term accounts for the change in specific entropy due to heat input, and the second accounts for introduction of new particles into the system.  

Supernovae are not the only heat source.  Mass ejected from old stars carries a specific kinetic energy per particle $\approx 3 \mu m_p \sigma_v^2 / 2$ that becomes thermalized in the ambient medium as the ejecta merge with it \citep[e.g.,][]{Conroy2014arXiv1406.3026C}.  The specific energy ejected by the old stellar population ($3kT_{H}/2$) therefore separates into a supernova heating term ($3kT_{\rm SN}/2$) and a kinetic energy term:
\begin{equation}
   \frac {3} {2} kT_{H} \equiv \frac {H \mu m_p} {\alpha \rho_*}
                \approx \frac {3} {2} \left[ kT_{\rm SN} + \mu m_p \sigma_v^2 \right] \; \; ,
\end{equation}
Substituting for $H$ in the entropy equation and multiplying by $rv^{-1}$ then leads to
\begin{equation}
  \frac {d \ln K} {d \ln r} \approx \left [ \frac {T_H} {T} - \frac {5} {3} \right] \frac {3 \rho_*} {\bar{\rho}_*}  \, \, ,
\end{equation}
where $\bar{\rho}_*$ is the mean stellar mass density within $r$.  In other words, the power-law slope of the radial entropy gradient depends primarily on the ratio of specific energy per ejected particle to thermal energy per ambient particle.  If that ratio exceeds 5/3, then the entropy gradient of the outflow will be positive, and if it does not, then the entropy gradient will be negative and convectively unstable.  Furthermore, the density factor can be neglected for the stellar mass distribution we have assumed, in which $\bar{\rho}_* \approx 3 \rho_*$. 

This relationship conveniently allows us to infer the supernova heating rate from the slope of the entropy gradient at 1--10~kpc. For the single-phase ellipticals, in which $K \propto r$, we obtain
\begin{equation}
  kT_{\rm SN} \, \approx \, \frac {8} {3} kT - \mu m_p \sigma_v^2 \, \approx \, 2 \, {\rm keV} 
\end{equation}
since they have $kT \approx 1.0 \, {\rm keV}$ and $\sigma_v \approx 300 \, {\rm km \, s^{-1}}$. In order to impart this much energy to the matter ejected by the old stellar population, the specific SNIa rate needs to be
\begin{equation}
    	\sim 0.3 \, {\rm (100 \, yr)^{-1} \, (10^{11} M_\odot )^{-1}} E_{51}^{-1} \alpha_{-19} \, \, ,
\end{equation}
where $E_{\rm SN} = (10^{51} \, {\rm erg}) E_{51}$ is the heat energy introduced per supernova into the ambient gas.  

This result is broadly consistent with direct observations of SNIa rates in old stellar populations \citep[e.g.,][]{Maoz2012MNRAS.426.3282M} and validates our outflow analysis.  However, it is in tension with the iron abundances observed in the hot gas, which are approximately solar.  An iron yield $\sim 0.7 \, M_\odot$ per supernova is equivalent to a mass fraction $\sim (7 \times 10^{-3}) E_{51}^{-1}$ relative to all the stellar ejecta, amounting to $\sim 5 E_{51}^{-1}$ times the solar abundance.  Reconciliation of the observed gas-phase iron abundances with the observed SNIa rate would therefore seem to require the newly-produced iron to be poorly mixed with the ambient gas, at least within the central 10~kpc of the outflow \citep[e.g.,][]{Tang+2009MNRAS.398.1468T}.

Now let us return to the issue of radiative cooling.  The cooling-free outflow solution for single-phase ellipticals must become invalid at small radii, because radiative cooling per unit volume scales as $n^2 \propto r^{-2.6}$ while heating scales as $\rho_* \propto r^{-2}$.  Defining $n_{e,{\rm SNbal}} \equiv H^{1/2} \Lambda^{-1/2}$ and $K_{\rm SNbal} \equiv kTn_{e,{\rm SNbal}}^{-2/3}$ leads to the following expression for the unstable locus of supernova-heating balance for a galaxy with $kT \approx 0.8 \, {\rm keV}$ and $\sigma_v \approx 300 \, {\rm km \, s^{-1}}$:
\begin{equation}
  K_{\rm SNbal} \approx (5 \, {\rm keV\, cm^2})  \, \alpha_{-19}^{-1/3} \, \Lambda_{3e-23}^{1/3} \,  r_{\rm kpc}^{2/3} \; \;   .
\end{equation}
This is shown with a green dashed line in the bottom panel of Figure~1.  The power-law profiles of density and entropy observed in single-phase ellipticals do indeed break where they intersect this locus, indicating that supernova heating exceeds radiative cooling at larger radii, and that supernova sweeping alone cannot rid the inner $\sim 1$~kpc of its stellar ejecta.

\section{Bondi Accretion \& Precipitation}
\label{sec-Bondi}

Inside of $\sim 0.5$~kpc, nine of the ten galaxies in our sample become isentropic, with core entropy $K_0 \approx 2 \,{\rm keV}$.  These all have radio power $\sim 10^{38} \, {\rm erg \, s^{-1}}$.  Adopting the conversion from radio power to jet power from \citet{Cavagnolo2010ApJ...720.1066C} indicates a typical jet power $\sim 2 \times 10^{42} \, {\rm erg \, s^{-1}}$.  The total power available from Bondi accretion \citep[see also][]{Allen+2006MNRAS.372...21A} can be estimated from the typical core entropy and black hole mass ($M_{\rm BH}$) for these galaxies:
\begin{eqnarray}
  \dot{M}_{\rm Bondi} c^2 & \sim & 4 \pi G^2 M_{\rm BH}^2 c^2 (\mu m_p)^{5/2} \left( \frac {5K_0} {3} \right)^{-3/2} \\
   & \sim & (5 \times 10^{44} \, {\rm erg \, s^{-1}})  \, M_{\rm BH,9}^2 \, K_2^{-3/2} \,\, ,
\end{eqnarray}
where $M_{\rm BH,9} \equiv M_{\rm BH} / 10^9 M_\odot$ and $K_2 \equiv K_0 / 2 \, {\rm keV \,cm^2}$.
Dividing the jet power by this value gives an inferred jet efficiency $\sim 10^{-2.4}$, meaning that the jets in these galaxies can plausibly be fueled by Bondi accretion onto the central black hole at the present time.

The tenth galaxy is NGC~4261, shown with purple stars in Figure 1.  It is the only one that drops below the precipitation threshold within $0.5 \, {\rm kpc}$ (see Figure 2), and its radio luminosity ($\sim 3 \times 10^{40} \, {\rm erg \, s^{-1}}$) corresponds to a jet power $\sim 10^{44} \, {\rm erg \, s^{-1}}$, almost two orders of magnitude greater than in the other galaxies.  From the central black-hole mass of this galaxy \citep[$5 \times 10^8 \, M_\odot$,][]{KormendyHo2013ARAA..51..511K} and its entropy in the innermost bin, we infer a total Bondi power $\sim 3.5 \times 10^{44} \, {\rm erg \, s^{-1}}$, which would imply an implausibly large jet efficiency $\sim 0.3$.  We therefore suggest that NGC~4261 may be a case in which cold chaotic accretion triggered by precipitation of cold clouds has temporarily boosted the accretion rate far above the Bondi level, as predicted by \citet{Gaspari+2013MNRAS.432.3401G,Gaspari+2014arXiv1407.7531G}.

\begin{figure}[t]
\includegraphics[width=3.5in, trim = 0.0in 0.0in 0.0in 0.0in]{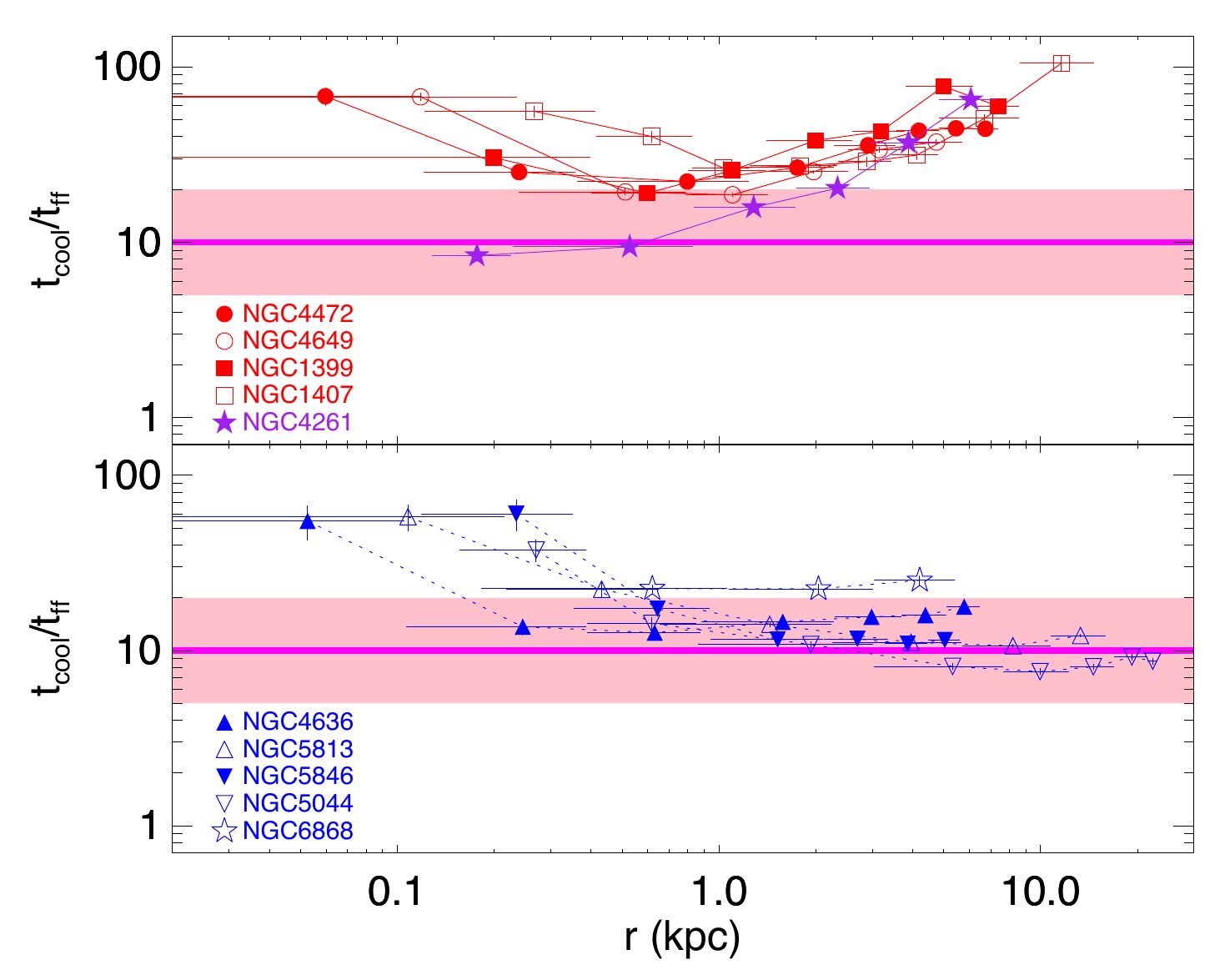} \\
\caption{ \footnotesize 
The precipitation criterion $\tc/\tff$ as a function of radius in single-phase ellipticals (top panel) and multiphase ellipticals (bottom panel).  Single-phase ellipticals all remain above the (pink) precipitation zone at $5 < \tc / \tff < 20$, with one exception:  NGC~4261 (purple stars) drops below the precipitation threshold at $\tc / \tff \approx 10$ (magenta line) within the central kpc.  In contrast, the multiphase ellipticals track the precipitation threshold in the 1--10~kpc region where  multiphase gas is found.  Notably, the jets from NGC~4261 are two orders of magnitude more powerful than those from the other nine galaxies in this sample, indicating that chaotic cold accretion of precipitating clouds, not Bondi accretion, is fueling the black-hole engine in this system.  
\vspace*{1em}
\label{k0-lha}}
\end{figure}

In all cases, the jet power greatly exceeds the $10^{41} \, {\rm erg \, s^{-1}}$ X-ray luminosity from $\lesssim 1$~kpc, implying that the jets are poorly coupled to the isentropic core and deposit most of their energy at larger radii.  This finding agrees with the radio morphologies of the jets, which extend well beyond the central kpc.  Tapping a small percentage of their power would compensate for cooling within $\sim 1$~kpc, and the flatness of the inner entropy profiles suggests that some combination of turbulent mixing and convection distributes that heat energy.  Furthermore, if the jet ever fails to couple to the isentropic core, the core will fill up with stellar ejecta, thereby lowering $K_0$ until the system reaches the precipitation limit.  According to models of cold chaotic accretion, the jet power should then rise by two orders of magnitude, as observed in NGC 4261.  In this sense, precipitation-driven feedback is like a backup system which ensures that the central cooling time remains $\gtrsim 10 \tff$, should either supernova sweeping or Bondi accretion ever fail to keep stellar ejecta from accumulating within the galaxy.

\section{Jet Coupling}
\label{sec-coupling}

A simple scaling argument suggests that the jets couple poorly to the inner regions of ellipticals because of their power.  In an isothermal potential, gas at the precipitation threshold has $K \propto r^{2/3}$ and $| d \ln T / d \ln r | \ll 1$, implying $n_e \propto r^{-1}$ and $kT \approx 2 \mu m_p \sigma_v^2 \approx 0.8 \, \sigma_{250}^2 \, {\rm keV}$.  Setting $\tc \approx 10 \tff$ therefore gives
\begin{eqnarray}
  K(r) & \, \approx \, & (3.6 \, {\rm keV \, cm^2}) \, \Lambda_{3e-23}^{2/3} r_{\rm kpc}^{2/3} \\
  n_e(r) & \, \approx \, & (0.1 \, {\rm cm^{-3}}) \, \Lambda_{3e-23}^{-1} \, \sigma_{250}^3 \, r_{\rm kpc}^{-1} \\
  L_{\rm X}(<r) & \approx & (10^{41} \, {\rm erg \, s^{-1}}) \, \Lambda_{3e-23}^{-1} \, \sigma_{250}^6 \, r_{\rm kpc}
\end{eqnarray}
where $L_{\rm X}$ is the X-ray luminosity.  Following \citet{vd05}, we estimate the shock velocity $v_{\rm sh}$ driven through gas with this density profile by a jet of power $P_{\rm jet} = (10^{42} \, {\rm erg \, s^{-1}}) P_{42}$ by setting $m_p n_e r^3 v_{\rm sh}^2 \sim r v_{\rm sh}^{-1} P_{\rm jet} $ and obtaining $v_{\rm sh} \sim (800 \, {\rm km \, s^{-1}}) P_{42}^{1/3} \Lambda_{3e-23}^{1/3} \sigma_{250}^{-1} r_{\rm kpc}^{-1/3}$.  For this shock velocity, the entropy jump condition from \citet{Voit+03} yields an entropy increment
\begin{eqnarray}
  \Delta K  & \, \approx \, & \frac {\mu m_p v_{\rm sh}^2} {3 (4 n_e)^{2/3}} \\ 
     & \, \sim \, & (2.4 \, {\rm keV \, cm^2}) \, P_{42}^{2/3} \, \Lambda_{3e-23}^{4/3} \, \sigma_{250}^{-4} \, \, \, \nonumber 
\end{eqnarray}
that is independent of radius.  

In other words, jet outbursts of the power observed in nine of these ten systems suffice only to boost $K_0$ to the observed level and have little effect on the $\sim 1$--10 kpc region.  That is because they deposit the bulk of their power at greater radii. Most of it is probably thermalized beyond the radius $r_{\rm th} \sim (10 \, {\rm kpc}) P_{42}  \sigma_{250}^{-2} \Lambda_{3e-23}$ at which jet-driven shocks become subsonic.  This is also the radius at which the typical jet power is comparable to the X-ray luminosity in the precipitating configuration, which is probably not a coincidence.  Furthermore, the excellent agreement of single-phase ellipticals with the power-law supernova sweeping model, along with the lack of scatter in their $K(r)$ profiles at $< 10$~kpc, supports the idea that most of their jet power is deposited farther out.  Even NGC~4261 agrees well with the others at 1--10 kpc, implying that the vast majority of its $\sim 10^{44} \, {\rm erg \, s^{-1}}$ jet power passes straight through the 1--10 kpc region with very little dissipation.

\section{The Black-Hole Feedback Valve}
\label{sec-valve}

We conclude with a hypothesis for the cessation of star formation known as quenching.  It would appear from the data presented in \S2 that black-hole feedback alone cannot completely quench massive elliptical galaxies because some have persistent multiphase gas.  Instead, massive elliptical galaxies separate into two groups.  In single-phase ellipticals, SNIa keep stellar ejecta from accumulating by sweeping it out of the central 1--10 kpc (\S3).  In multiphase ellipticals, supernova heating cannot quite overcome radiative cooling in the 1--10 kpc region.  These galaxies therefore fall into a precipitating state in which black-hole feedback maintains $\tc \approx 10 \tff$ and can cannot entirely shut off star formation (\S4).  Jet power can fluctuate between $\sim 10^{42} \, {\rm erg \, s^{-1}}$ (Bondi-like accretion) and $\sim 10^{44} \, {\rm erg \, s^{-1}}$ (cold chaotic accretion) without disrupting the 1--10~kpc region because the jets deposit most of their energy beyond $\sim 10$~kpc (\S5).

Why then are the five ellipticals with $\sigma_v > 260 \, {\rm km \, s^{-1}}$ fully quenched while low-level star formation persists in the five multiphase ellipticals with lower velocity dispersions?  Some insight can be gained by comparing $K_{\rm precip}$ with $K_{\rm SNbal}$.  In galaxies with $K_{\rm precip} > K_{\rm SNbal}$, the interstellar medium can be thermally unstable even if supernova heating exceeds radiative cooling. We therefore expect those galaxies to experience multiphase circulation in which SNIa drive hot gas out of the central regions where the stars are but cool clouds precipitate back in from the margins.  In contrast, galaxies with $K_{\rm SNbal} > K_{\rm precip}$ can drive outflows capable of shutting off precipitation, by raising $\tc \approx ( 5 \, {\rm Gyr} ) T_{\rm keV}^{-1/2} \Lambda_{3e-23} (K / 100 \, {\rm keV \, cm^2})^{3/2}$ high enough to prevent the outflowing gas from cooling after it leaves the galaxy.

The condition for avoiding potentially star-forming multiphase gas is then
\begin{equation}
  \frac {K_{\rm precip}} {K_{\rm SNbal}} \approx 0.6 \, \alpha_{-19}^{1/3} \, \Lambda_{3e-23}^{1/3} \, \sigma_{250}^{-4/3} 
     \, < 1 \; \; , 
\end{equation}
implying that galaxies with $\sigma_v \lesssim (200 \, {\rm km \, s^{-1}} ) \alpha_{-19}^{1/4} $ $\Lambda_{3e-23}^{1/4}$ should be embedded within a precipitating circumgalactic medium.  Intriguingly, this result implies that quenching depends more directly on the concentration of stellar mass toward the center of a galaxy than on the total stellar mass itself, in qualitative agreement with observations \citep[e.g.,][]{FangFaber2013ApJ...776...63F}.  It also implies that quenching requires a greater value of $\sigma_v$ at earlier times, when the stellar population is younger and $\alpha$ is greater.  However, the bottom panel of Figure~1 implies that something more is needed for quenching to be complete.

We hypothesize that black-hole feedback outbursts are also required to flip massive ellipticals with $K_{\rm SNbal} > K_{\rm precip}$ from the precipitation locus to a supernova-sweeping state.  This ``black-hole feedback valve" can accomplish the task with an outburst powerful enough to produce an entropy increase $\Delta K > K_{\rm SNbal} - K_{\rm precip}$.  It will be interesting to test this quenching mechanism and see how it plays out in simulated elliptical galaxies. 

\vspace*{1.0em}

The authors thank J. Bregman, A. Crocker, M. Gaspari, and B. McNamara for helpful conversations.  GMV and MD acknowledge NSF for support through grant AST-0908819.  G.L.B. acknowledges NSF AST-1008134, AST-1210890, NASA grant NNX12AH41G, and XSEDE Computational resources.

\bibliographystyle{apj}

\end{document}